# On the use of Water and Methanol with Zeolites for Heat Transfer

Rafael M. Madero-Castro[a], Azahara Luna-Triguero[b,c], Andrzej Sławek[d,e], José Manuel Vicent-Luna[f]*, and Sofia Calero[a,f]*

[a]Department of Physical, Chemical, and Natural Systems, Universidad Pablo de Olavide. Ctra. Utrera km. 1. ES-41013 Seville, Spain.
[b]Energy Technology, Department of Mechanical Engineering, Eindhoven University of Technology, P.O. Box 513, 5600 MB Eindhoven, The Netherlands.
[c]Eindhoven Institute for Renewable Energy Systems (EIRES), Eindhoven University of Technology, PO Box 513, Eindhoven 5600 MB, the Netherlands
[d]Academic Centre for Materials and Nanotechnology, AGH University of Science and Technology, Kawiory 30, 30-055 Kraków, Poland.
[e]Faculty of Chemistry, Jagiellonian University, Gronostajowa 2, 30-387 Kraków, Poland.
[f]Materials Simulation and Modelling, Department of Applied Physics, Eindhoven University of Technology, P.O. Box 513, 5600MB Eindhoven, The Netherlands.

**ABSTRACT**

Reducing carbon dioxide emissions has become a must in society, being crucial to find alternatives to supply the energy demand. Adsorption-based cooling and heating technologies are receiving attention for thermal energy storage applications. In this paper, we study the adsorption of polar working fluids in hydrophobic and hydrophilic zeolites by means of experimental quasi-equilibrated temperature-programmed desorption and adsorption combined with Monte Carlo simulations. We measured and computed water and methanol adsorption isobars in high-silica HS-FAU, NaY, and NaX zeolites. We use the experimental adsorption isobars to develop a set of parameters to model the interaction between methanol and the zeolite and cations. Once having the adsorption of these polar molecules, we use a mathematical model based on the potential theory of adsorption of Dubinin-Polanyi to assess the performance of the adsorbate-working fluids for heat storage applications. We found that the use of molecular simulation is an alternative for investigating energy storage applications since we can reproduce, complement, and extend experimental observations. Our results highlight the importance of controlling the hydrophilic/hydrophobic nature of the zeolites by changing the Al content to maximize the working conditions of the heat storage device.

## 1. INTRODUCTION

The decrease in energy consumption is essential for the mitigation of global warming.[1–3] The use of renewable energies along with the reduction of fossil fuels is important here.[4,5] To mitigate this problem, solar, wind energy, or biofuels are promising candidates, but the intermittent nature of renewable energies limits their application for society consumption. This is motivating researchers to work on new approaches for the storage of renewable energy.[6–8]

There are many methods for energy storage at industrial level based on converting renewable energy into potential energy. One of the most used is Pumped-Storage Hydroelectricity (PSH) or Pumped Hydro Energy Storage (PHES). PSH uses the surplus energy obtained in hydroelectric dams in low power demand hours to elevate water from lower to higher levels. That converts the surplus energy into potential energy, which can be used in high power demand periods.[9–15] However, hydroelectric dams have an impact on the environment. .[16] Another method is the Compressed Air Energy Storage (CAES).[6,17–20] CAES uses renewable energies, mainly wind, to compress air at high pressure and generate electricity.[21] Based on the same principle of stored mechanical energy, Flywheel Energy Storage (FES) uses inertia for the storage.[22] The operation of the system consists of a rotor that is driven and keeps spinning to store kinetic energy.[19,23–26] The best known storage method is electrochemical storage, specifically lithium-ion batteries.[27–30] The current expansion of cell phones or hybrid cars [31] has increased the need of developing the technological market in this field.[32] The main limitations of batteries are the loss of capacity [33] and the risk of thermal runaways or explosions.[29]

In the context of finding alternative methods, Thermal Energy Storage (TES) in three variants (using sensible, latent, or thermochemical heat) has been proposed.[8] This technology is based on the adsorption of (gas-liquid) adsorbates with solid adsorbents, releasing energy during the process. Thus, the efficiency of the heat storage process strongly depends on the adsorbent-adsorbate interactions. Porous materials such as Metal-Organic Frameworks (MOFs),[34–36] silica gels,[37–39] activated carbons,[40–42] or zeolites,[43–46] have proven to be promising candidates for this application. The most common adsorbates are water,[47,48] ammonia [34,49] and light alcohols,[50,51] being water the most studied molecule for storing energy in diverse porous materials.[51–53]

Because adsorption-based energy storage is a promising alternative, the number of experimental and simulation studies is rising. In this context, numerical modeling and molecular simulations are excellent tools to complement experiments. Tatsidjodoung *et al.*[44] studied the water-NaX zeolite pair to store thermal energy from the sun. This work concluded that, although there are slight discrepancies between experiments and numerical calculations, simulation is an excellent method for making feasible predictions. Semprini *et al.*[52] studied the energy transfer between the 13XBF zeolite and water and its orientation towards the construction of refrigerants, finding a good agreement between simulations and experiments. Lehmann *et al.*[53] studied the influence of the cation (sodium or calcium) in zeolite X and water working pairs for energy storage applications. They revealed the importance of working conditions, such as vapor pressure, in the thermochemical energy properties such as the Storage Density (SD). Similarly, Kohler *et al.*[51] studied the energy stored in zeolite NaX using water as a working fluid, showing the influence of the desorption temperature in the storage density. They compared their values with the energy stored by activated carbons with alcohols as working fluids and noted that the adsorption capacity is as important as the interactions with the adsorbent. Stach *et al.*[54] studied the influence of Na and Mg cations and their ratio in zeolites and silica gels using water. Most studies in the literature involve NaX zeolite (with FAU topology) as it is one of the most popular commercial zeolites. It is worth mentioning that NaX usually operates at very high desorption temperatures, typically over 500 K. This is due to the high hydrophilicity of the structure caused by the content of sodium cations. However, other FAU-type zeolites are proposed as interesting alternatives. Ristić *et al.*[48] highlighted the significance of decreasing the desorption temperature to optimize the low-temperature heat storage density. To this end, they used NaY zeolite, which is equivalent to NaX but with a slightly higher Si/Al ratio, to study the adsorption heat storage with water as working fluid. To reduce the desorption temperature of the water they proposed six post-synthesis modified samples from chemical treatment of NaY. The modified adsorbents reduced the desorption temperature up to 30 K, showing a maximum performance at temperatures *circa* 400 K. This is an improvement compared to the operating conditions of the NaX zeolite as discussed above. However, for low-temperature applications a working pair that lower the desorption temperature near room conditions is preferred. In this regard, we propose to regulate the hydrophilic degree of the adsorbent by controlling the Si/Al ratio of the zeolite. This way, the performance of a low-temperature process can be maximized while avoiding the post-synthesis treatment step, which may reduce the production costs.

This work combines experimental techniques, molecular simulation, and thermodynamical modeling for the study of water and methanol adsorption-based energy storage in FAU-type zeolites (FAU)[55,56] with different Si/Al ratio.[57] We analyze the effect of the hydrophobic/hydrophilic nature of the adsorbent in the adsorption behavior, external operating conditions, and energy storage. We used quasi-equilibrated temperature-programmed desorption and adsorption (QE-TPDA) experiments to measure the adsorption isobars of the working pairs. Molecular simulation was used to shed light on the adsorption mechanism from the atomistic level. To this aim, we developed a set of Lennard-Jones parameters that defines the FAU-methanol interactions independently of the ratio of cations in high-silica (HS) FAU, NaY, and NaX structures. Finally, we are using a thermodynamical model to correlate the adsorption properties with the energy storage of each particular working fluid-zeolite pair.

## 2. METHODOLOGY

### 2.1. Experimental details.

Three samples of FAU were used for the adsorption experiments. HS-FAU is $Na^+$ exchanged dealuminated high-silica faujasite with Si/Al > 100, NaY is $Na^+$ exchanged faujasite with Si/Al ≈ 2.61, while NaX is $Na^+$ exchanged faujasite with Si/Al ≈ 1.06. The characteristics of these materials, *i.e.,* low-temperature nitrogen adsorption and powder X-ray diffraction, were reported in our previous works.[58,59]

Adsorption measurements were performed using quasi-equilibrated temperature-programmed desorption and adsorption (QE-TPDA) technique. This instrument is a homemade modified setup for temperature-programed desorption (TPD), which was described in detail in previous works.[60,61] The samples of 7–10 mg were placed in a quartz tube and activated by heating in He flow (6.75 cm$^3$ min$^{-1}$) up to 400 °C (HS-FAU, NaY) or 500 °C (NaX) with a 10 °C min$^{-1}$ ramp and cooling it down to RT. Adsorption was measured in flow of He containing a water steam (saturated) or methanol vapors ($p/p_0$ < 0.05). The samples were heated to induce desorption and cooled to induce adsorption. Each profile was averaged over 3 desorption–adsorption cycles. For methanol we used 4 °C min$^{-1}$ ramp for all materials, while for water 2 °C min$^{-1}$ for NaY and NaX and 1 °C min$^{-1}$ for HS-FAU. Between each cycle, they were kept at RT for 2 hours. More details on data reduction and methodology are available in the literature.[62]

### 2.2. Simulation details.

We carried out Monte Carlo simulation in the Grand Canonical ensemble (GCMC) to obtain the adsorption properties of water and methanol in the three selected zeolites. We performed a minimum of 5·10$^5$ MC cycles to ensure the adsorption data is fluctuating around equilibrium values. After the equilibration procedure, we conducted additional 2·10$^5$ cycles for the final production runs. All simulations were performed using the RASPA simulation software.[63] Adsorbent-adsorbate, adsorbate-adsorbate, and adsorbate-cation interactions were defined with van der Waals and electrostatic interactions via the Lennard-Jones and Coulombic potentials, respectively, while we use a Coulombic potential to model the adsorbent-cation interaction. We truncated the potential with an effective cut off of 12 Å and we used the Ewald summation method [64] to compute the long range electrostatic interactions.

The adsorbents are zeolites with FAU topology; NaX, NaY, and HS-FAU with Si/Al ratio of 1.06, 2.61, and 100, respectively. These structures contain 88, 56, and 2 Al in the unit cell, respectively, and the same number of Na cations to compensate the net negative charge of the system. The structural models were reported in previous works,[58,59,65,66] which were created following the methodology developed by Balestra *et al.*[67]: (i) random distribution of Si atoms following Lowenstein's rules, (ii) extra framework-cationsinitially located at their crystallographic positions, and (iii) structural minimization using Baker's [68] method with full-flexible core-shell potential.[69,70] More details about the assembly of the structures can be found in the Supporting Information (Section S1).

To describe the molecules of water we used the flexible SPC/E model [71,72] and for methanol, the TraPPE model.[73] Force fields to model the water and methanol adsorption curves in zeolites can be found in literature. [74–77] Xiong, R. [75] studied the interaction between molecules of water and alcohol with pure silica type MFI-zeolite, but without extra framework cations in the system. Di Lella *et al.*[74] provided a set of parameters to reproduce water adsorption in FAU-topology zeolites. However, the parameters and charges of each zeolite-water pair are dependent of the composition of the adsorbent, making this set highly specific and non-transferable. In this work, we used a transferable set of Lennard-Jones parameters and zeolite and cation point charges [78] that are independent of the Si/Al ratio. Specific Lennard-Jones parameters for the pair interactions for water-zeolite were taken from our previous work,[59] while for methanol-zeolite they were unavailable. To sort this out, we parameterized these host-guest interactions by fitting to the experimental adsorption isobars measured with QE-TPDA experiments. Additional details about the parameterization procedure and final set parameters can be found in the Supporting Information (Section S2 and Tables S1 and S2). For the crossed interactions, we used Lorentz-Berthelot mixing rules.[79] The sodalites cages in FAU zeolites, accessible to water molecules, were artificially blocked for the molecules of methanol.

**2.3. Thermodynamical model.**

QE-TPDA experiments and GCMC simulations provided the adsorption properties of zeolites-fluids working pairs. Using these results and a mathematical model based on the adsorption theory of Dubinin-Polanyi [80] we predicted the adsorption-based energy storage. The complete description of the thermodynamical model can be found in the Supporting Information and is also available in the literature.[81] In short, we first used the adsorption isobars and isotherms to calculate the adsorption characteristic curve.[82] This reduces the two-dimensional relation between loading ($q(p,T)$), temperature ($T$), and pressure ($p$), to the temperature-pressure invariant characteristic curve ($W(A)$). This curve describes the relation between the specific volume of the adsorbed fluid ($W$) with the adsorption potential ($A$) or Gibbs free energy. Using the characteristic curve, we can determine the loading dependence of the adsorption enthalpy ($\Delta H$), which depends on the vaporization enthalpy of the fluid ($\Delta H_{vap}$), the slope of the characteristic curve, and the entropy changes ($\Delta S$). Finally, we obtained the thermochemical storage density ($SD$) of each working fluid from the numerical integration of the adsorption enthalpy as a function of the loading within the adsorption and desorption temperature range (see Section S3 of the Supporting Information for specific details).

## 3. RESULTS AND DISCUSSION

Figure 1 shows the QE-TPDA profiles in the studied faujasites, where the profile above the baseline (ssr = 0) reflects the desorption process and the profile below the baseline reflects the adsorption process. The intensity of the profiles corresponds to the instantaneous concentration of adsorbate desorbed or adsorbed in the material at a given temperature.

The profiles reveal differences in the adsorption of water and methanol. For HS-FAU we found very sharp profiles both for water and methanol. This means that adsorption occurs abruptly in a narrow temperature range. For NaY and NaX, the low-temperature adsorption at 300–350 K corresponds to high density states where the guest-guest interactions are of great importance. Figure 1a shows that most water is adsorbed in NaY and NaX between 375−475 K. A long tail at higher temperatures is most likely due to the interactions of the water molecules with the cations. This effect is more pronounced for NaX, which has more cations than NaY. The profiles for methanol (Figure 1b) are similar than for water. Desorption maxima and adsorption minima for NaY and NaX are shifted towards higher temperatures than for water, up to *circa* 530 K. Also, the broad high-temperature tail for NaX is extraordinarily intensive. Generally, the QE-TPDA profiles show that adsorption is stronger for methanol than for water in NaY and NaX. T interactions between methanol and NaX cations are particularly strong.

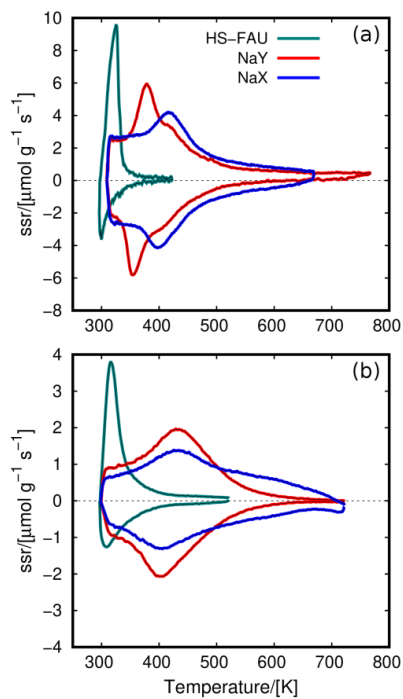

**Figure 1.** QE-TPDA profiles of (a) water and (b) methanol in FAU zeolites. The values of partial

pressure for water and methanol in HS-FAU, NaY, and NaX are 1.98, 2.8, and 3.1 kPa, respectively (water) and 0.7 kPa (methanol).

The adsorption isobars can be obtained by integrating the QE-TPDA profiles.[62] We used the adsorption isobars of methanol for the parameterization of the force field required for molecular simulation (Table S2). Figure 2 compares the experimental and computed adsorption isobars under the same working conditions (see Figure 1). Considering that we are using the same set of (transferable) parameters and partial charges for all the systems, we found good agreement with the experimental results.

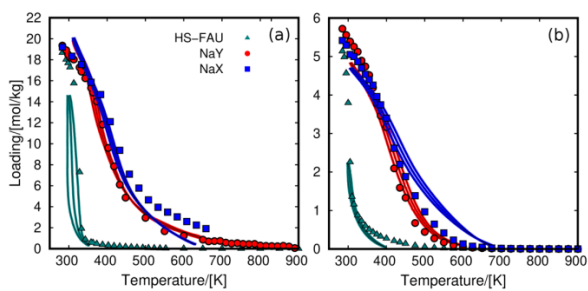

**Figure 2**. Experimental (lines) and calculated (symbols) adsorption isobars for (a) water and (b) methanol in FAU zeolites. Each experimental isobar is divided into three curves corresponding to adsorption and desorption cycles (obtained from the QE-TPDA profiles) and the average of them. The values of partial pressure for water and methanol in HS-FAU, NaY, and NaX are 1.98, 2.8, and 3.1 kPa, respectively (water) and 0.7 kPa (methanol).

The behavior of the adsorption isobars is similar for water and methanol, since both fluids are polar. The hydrophobicity of HS-FAU, due to the low content of Na cations, leads to a steeped isobar at low values of temperature. As increasing the cations content in the zeolites, the shape of the isobar shows a smooth loading decrease reaching desorption temperatures at about 600 K. This proves the high affinity of polar fluids for the extra framework cations of the zeolites. The adsorption of methanol in the Na-FAU zeolites shows a minor hysteresis loop, a displacement between adsorption and desorption. Similarly, the adsorption isobar of water in HS-FAU shows a tiny hysteresis loop, which is lower for the zeolites with higher cation content. The set of parameters was then fitted to the intermediate curve, which is the average of adsorption/desorption from experimental measurements (Figure 2).

Since partial pressures of methanol and water adsorption are different, it is not possible to directly compare saturation loadings of the two fluids from the adsorption isobars. However, we can convert each adsorption isobar to their corresponding characteristic curve, which only depends on the fluid-zeolite working pairs. Figure 3 shows the characteristic curves of water and methanol obtained from the adsorption data. For the GCMC curve, we used the adsorption isobar from Figure 2 and additional adsorption isotherms (Figure S1) to complete the characteristic curve, ranging from zero-coverage to saturation conditions. The data from independent adsorption isobars and isotherms fall into the same characteristic curve. We found that the volumetric adsorption is considerably higher for water (about 0.35 ml of fluid per gram of adsorbent) than for methanol (0.25 ml/g). This is due to the smaller size of water that can connect through four hydrogen bonds per molecule. Methanol can connect through two,[83,84] leading to a worse molecular packaging. Another relevant factor for the higher adsorption of water compared to methanol is that, contrary to methanol, the water molecules can enter the small sodalite cages of FAU zeolites. For this reason, the free volume for the adsorbents is larger for water than for methanol. To increase the limited number of points obtained from the GCMC simulation, we use splines. It is important to use smooth functions that fit the data well to minimize the noise in the calculations involving the characteristic curves. The fitting for the experimental characteristic curve is more straightforward since it contains more points resulting from the measurements for small temperature increments. The characteristic curves were complemented with adsorption at high temperatures to reach the low coverage regime.

The presence of cations in the FAU zeolites does not alter their pore volume significantly.[58] This is why adsorption isobars and characteristic curves have similar saturation values, independently of the cation content. However, the concentration of cations influences the hydrophilic/hydrophobic nature of the zeolites. The adsorption trend of both fluids in NaY and NaX is very similar despite the differences in the number of cations. The curves in NaY are slightly shifted to lower values of temperature (Figure 2) or lower adsorption potential (Figure 3). This effect is not as strong as for the n-alkanes.[58] This suggests that the presence of cations in the zeolite is more important than the concentration. As discussed earlier, Ristić *et al.*[48] proposed a post-synthesis chemical treatment of NaY to control the desorption temperatures of water. However, all modified samples contained a similar Si/Al ratio, and they found a decrease of the desorption temperatures of about 30 K compared to the original NaY zeolite. To control the desorption temperatures over a wider range of working conditions, we suggest to reduce the cation content to values between HS-FAU (Si/Al ratio = 100) and NaY (Si/Al ratio = 2.61).

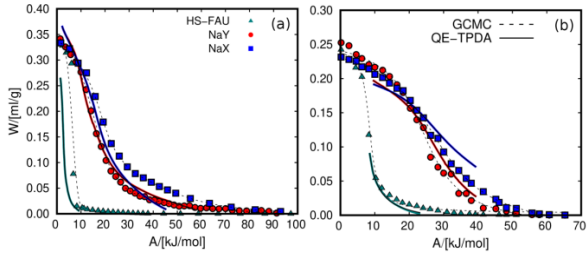

**Figure 3**. Characteristic curves of (a) water and (b) methanol in FAU zeolites using data from QE-TPDA (solid lines) and CGMC (dashed lines). The dashed lines represent the fitted curve using splines.

Reported adsorption studies for heat storage applications typically measure adsorption isotherms at a wide range of temperatures. Then, the adsorption isotherms are reduced to a common characteristic curve. Instead of doing this, here we measure and compute a single adsorption isobar to obtain the temperature dependence of the loading needed for further calculations of the storage density. Then, if necessary, we completed the low coverage regime of the computed curve with data from additional adsorption isotherms. To validate this approach, we compared the characteristic curves for water obtained in this work from QE-TPDA and GCMC simulation with those reported by Lehmann *et al.*[53] and Stach *et al.*[54] (Figure 4). Our results are in line with those reported in the literature, with slight deviations mainly due to the use of different commercial samples. In all cases, we observe that the saturation loading (corresponding to A→0 kJ/mol) converges to similar values, *i.e.*, about 0.35 ml/g, which is the saturation value for water in all FAU zeolites (Figure 3). The results shown in Figures 3 and 4 reveal the invariance of the characteristic curves with the adsorption conditions, thus giving consistency to the use of the DP theory for the working pairs of this work.

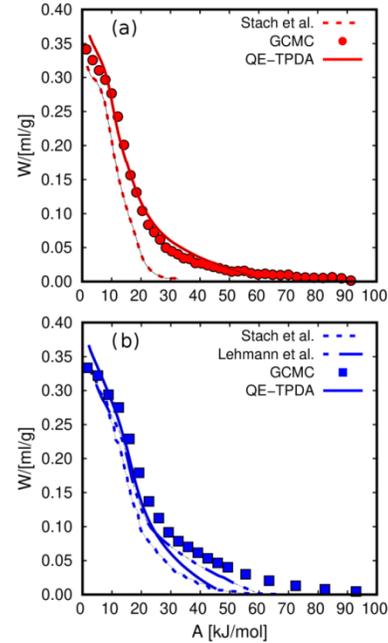

**Figure 4**. Characteristic curves of water in (a) NaY and (b) NaX obtained from GCMC (symbols) and experiments (lines). The experimental values are from QE-TPDA and reported by Stach *et al.*[54] and Lehmann *et al.*[53]

The performance of a working pair for adsorption-based heat storage depends on two thermodynamical quantities; the adsorption capacity and the adsorption enthalpy at the working conditions. However, these two quantities are not independent, and the adsorption enthalpy can be obtained from the adsorption data and the physicochemical properties of the working fluid. We take the data from the characteristic curves (Figure 3) to obtain the adsorption enthalpy of water and methanol in the three zeolites (Figures 5 and 6) using the DP theory as described in the methodology. The results obtained from the QE-TPDA experiments and GCMC simulation, depicted in Figures 5 and 6, are in agreement, showing similar differences to those found in Figures 2 and 3. Differences between the measured and the computed adsorption isobars shown in Figure 3 entail a deviation of less than 3 kJ/mol in the adsorption enthalpy, except for methanol in NaX, where the differences are about 10 kJ/mol.

The adsorption enthalpy depends on the adsorption behavior and on the physicochemical properties of the working fluid. The properties used in the DP formulation are the vaporization enthalpy, thermal expansion coefficients, liquid density, and saturation pressure. Although Figures 2 and 3 indicate similar behavior for water and methanol adsorption, we found variations in the adsorption enthalpy for these two fluids. These discrepancies are related to the physicochemical properties of water and methanol. Figure 5 shows the loading dependence on the adsorption enthalpy of water in the three zeolites. There is correlation between the number of cations (degree of hydrophilicity) and the adsorption enthalpy. At low coverage, the absolute value of the adsorption enthalpy is about 80 kJ/mol for the three structures, but the behavior is differentiable at intermediate and higher loadings. For HS-FAU, the adsorption enthalpy shows an abrupt decrease with loading after the low coverage regime. This phenomenon is related to the low concentration of cations that act as strong interaction centers. Once the first molecules of water are adsorbed near the cations at low coverage regime, they quickly nucleate and occupy the rest of the adsorption sites in the structure. The decrease in adsorption enthalpy in NaY and NaX is less pronounced than in HS-FAU due to the higher sodium content. At saturation, the adsorption enthalpy is about 50-55 kJ/mol because adsorbate-adsorbate interactions prevail over the interactions with the zeolite.

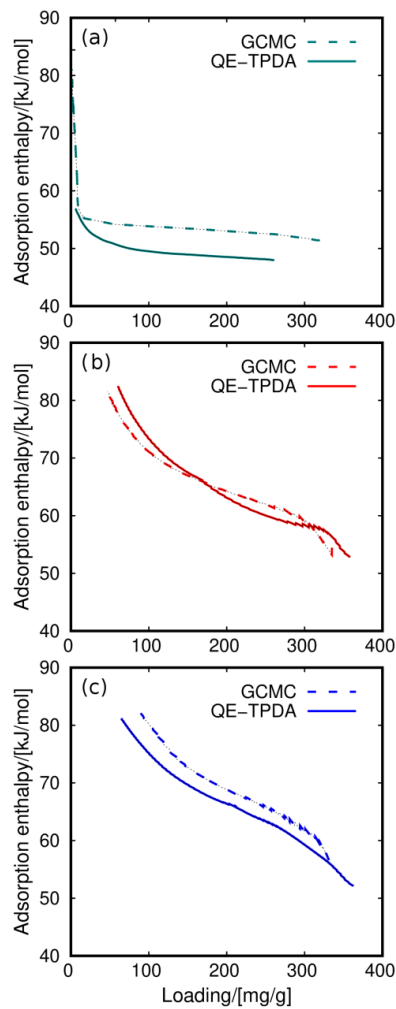

**Figure 5**. Adsorption enthalpy of water as a function of loading in (a) HS-FAU, (b) NaY, and (c) NaX. The values were obtained from GCMC simulation (dashed lines) and QE-TPDA (solid lines) data.

Figure 6 shows the adsorption enthalpy using methanol as working fluid. The general trend differs from the values for water shown in Figure 5. The curve corresponding to the adsorption of methanol in HS-FAU is like to that found for water. The values reach 80 kJ/mol at low coverage and immediately decrease to 50 kJ/mol. However, the sudden decrease of adsorption enthalpy at low coverage is less pronounced for methanol than for water, and the trend shifts slightly at high loading. The most remarkable differences are for NaY and NaX. At low coverage, the values are about 80 kJ/mol as for the other systems. At intermediate loading, the curves show a minimum at about 60-

65 kJ/mol and the adsorption enthalpy increases to 70-75 kJ/mol at high loading.

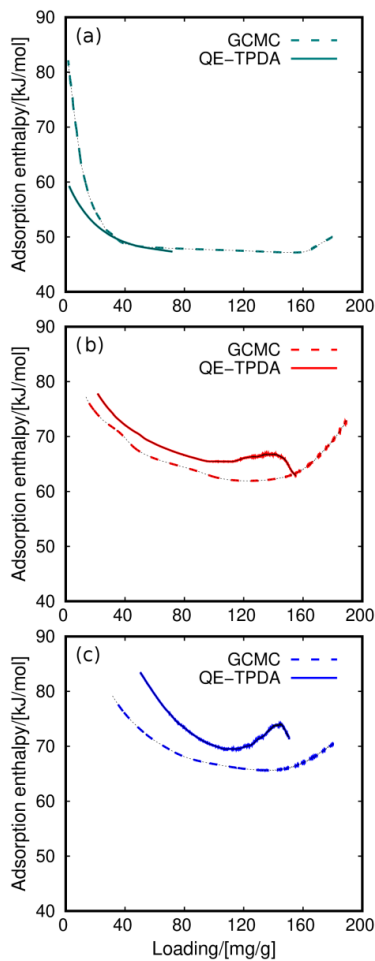

**Figure 6**. Adsorption enthalpy for methanol as a function of loading in (a) HS-FAU, (b) NaY, and (c) NaX. The values were obtained from GCMC simulation (dashed lines) and QE-TPDA (solid lines) data.

One of the main features of an adsorption heat storage device is its energy storage density or simply storage density. We calculated this quantity by integrating the adsorption enthalpy curves between fixed adsorption and desorption temperatures. The values shown in Figures 7a-c were obtained for a fixed adsorption temperature of 315 K. We select this temperature for being the lowest temperature measured in the QE-TPDA experiments for the three zeolites. The figures show the storage density as a function of the desorption temperature. From these figures it is possible to compare the values obtained with QE-TPDA and GCMC. These results show similar differences as in previous adsorption isobars (Figure 2) and adsorption enthalpy (Figure 5). However, the operating conditions play an important role here, making comparison more difficult. For example, Figure 7a shows the GCMC values for water in HS-FAU. Because of the high hydrophobicity of this zeolite, the adsorption obtained from QE-TPDA could not reach the saturation capacity of water in HS-FAU. This is because from the experimental side, establishing adsorption equilibrium in hydrophobic adsorbents takes long time. The driving force is very low, leading to condensation within the micropores. The inlet figure compares the storage density obtained from QE-TPDA and GCMC using adsorption temperatures of 300 K and 324 K, respectively. Using this approximation, the two curves show analogous abrupt increase, reaching similar storage density values. Extending the GCMC simulation to a wider range of temperature provides more detailed analysis of the storage density behavior. Therefore, it could lead to the optimization of the process based on the operational conditions for each working pair. To compare the maximum performance of the three adsorbents, we use the data from GCMC simulation and decrease the adsorption temperature to 300 K. This ensures that all systems reach saturation (Figure 7d). We found two trends; (i) HS-FAU shows an abrupt increase in the storage density. The maximum energy is released at relatively low temperatures compared to NaY and NaX because of the rapid desorption in this hydrophobic structure. For example, at a desorption temperature of 350 K, the storage density of HS-FAU surpasses 900 kJ/kg. At the same temperature, NaY and NaX do not even reach 600 kJ/kg. (ii) NaY and NaX have a moderate steep increase, reaching the maximum values at the higher tested desorption temperature, *i.e.*, 500K. NaY and NaX do not converge to the same storage density value because these zeolites have not released all the adsorbed water at 500 K (see Figure 2). In contrast, for HS-FAU, the curve is flat at temperature values above 350 K because the zeolite desorbs most of the molecules around this temperature. HS-FAU shows a maximum value of storage density lower than 1000 kJ/kg, while for NaY and NaX the value is higher than 1100 kJ/kg. It is important to mention that the working pressure of water adsorption in HS-FAU was set *circa* 1 kPa lower than for NaY and NaX (1.98 kPa for HS-FAU, 2.8 kPa for NaY, and 3.1 kPa for NaX). However, the lower value of pressure does not explain the low storage density obtained for HS-FAU (compared to NaY and NaX), since for these values of pressure the three zeolites adsorb similar amount of water at room temperature (Figure 2). The maximum storage density depends on the maximum loading of water that the adsorbent can capture and release and the exchange adsorption enthalpy. This means that the differences in the maximum storage densities showed in Figure 7 d) are mainly due to the adsorption enthalpy of water (see Figure 5).

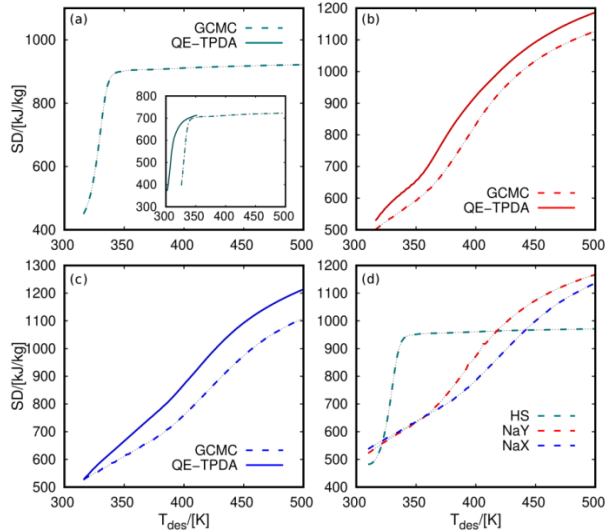

**Figure 7**. Storage density (SD) of water-zeolite pairs in (a) HS-FAU, (b) NaY, and (c) NaX at $T_{ads}$=315 K and P=1.98 kPa (HS-FAU), P=2.8 kPa (NaY), and P=3.1 kPa (NaX). The inlet figure in (a) is for $T_{ads}$=300 K (QE-TPDA) and $T_{ads}$=324K (GCMC). Figure (d) shows the storage density from GCMC simulation in three zeolites at $T_{ads}$=300 K.

Previous results highlight the importance of the operating conditions in order to maximize the performance of each fluid-adsorbent working pair. Many works using process simulations or experimental measurements compare the values of several working pairs at single fixed operating conditions. However, the storage density values could change drastically by slightly changing the operating temperature. To compare our approach with reported data, we computed the storage density of water in NaY and NaX at the same conditions used in previous studies. Ristić *et al*.[48] reported storage density of water in NaY of about 675 kJ/kg (187.5 Wh/kg) for fixed adsorption and desorption temperatures of 313 K and 413 K, respectively, and operating pressure of 1.23 kPa. Lehmann *et al*.[43] provided storage density values for water in NaX of about 815 kJ/kg (226.38 Wh/kg). However, in this case, the adsorption and desorption temperatures were extended to 293 K and 453 K, respectively. In principle, these two values cannot directly be compared, and one could think that NaX shows higher storage densities that NaY. However, extending the desorption temperature it is possible to analyze the performance of the two systems. In this regard, Figure 8 shows the computed storage density of water in NaY and NaX using reported adsorption conditions as a function of desorption temperature. For comparison, the figure also includes available experimental data. Our predictions are in agreement with the experiments and allow solid comparison between the performance of the two zeolites and water working pairs.

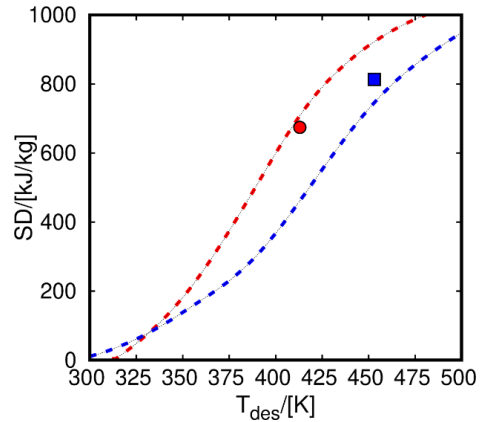

**Figure 8.** Storage density (SD) of water in NaY (red) and NaX (blue). The values taken from literature are indicated with symbols.[43,48] The values resulting from GCMC simulation are in dashed lines. The operational conditions are Tads=313 K and P=1.23 kPa for NaY and $T_{ads}$=293 K and P=3 kPa for NaX.

To compare the performance of the two working fluids, we calculated the storage density of methanol in the three zeolites using the data obtained with QE-TPDA and GCMC (Figure 9). As for water, differences between the two techniques are based on the differences found in the adsorption isobars (Figure 2) and the adsorption enthalpy (Figure 5).

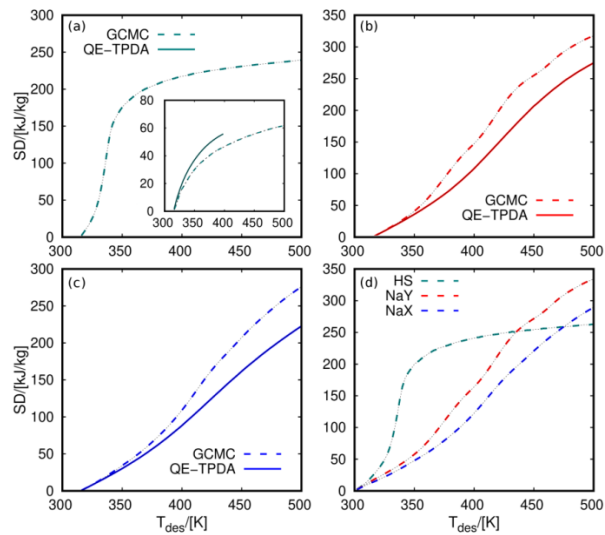

**Figure 9**. Storage density (SD) of methanol-zeolite pairs in (a) HS-FAU, (b) NaY, and (c) NaX at $T_{ads}$=315 K and P=5 kPa (HS-FAU) and P=0.7 kPa (NaY and NaX). The inlet figure in (a) is for $T_{ads}$=315 K and P=0.7 kPa. Figure (d) shows the storage density from GCMC simulation at $T_{ads}$=300 K and P=5 kPa (HS-FAU) and P=0.7 kPa (NaY and NaX).

The storage densities of methanol depicted in Figure 9 show the same trend than for water (Figure 7), but the maximum values are, on average, between 3 and 4 times lower. Despite the different trends on the adsorption enthalpy of water and methanol (Figures 5 and 6), the absolute values are similar. Another factor that could influence the performance of storage densities when comparing distinct fluids is the vaporization enthalpy. At room temperature, this enthalpy is about 10 kJ/mol higher for water than for methanol. However, the limiting factor comparing the storage densities of the two fluids is the difference of adsorption loading. Figure 2 shows that the FAU zeolites adsorb between 3 and 4 times more water than methanol at low temperatures, which is in line with the trend observed in the storage densities.

## 4. CONCLUSIONS

The combination of QE-TPDA experiments with MC simulation gives detailed information on the use of FAU zeolites for heat storage application. The calculated adsorption isobars show strong influence of the hydrophobic degree of the adsorbent in the desorption temperatures. HS-FAU desorbs most water and methanol at much lower temperature than NaY and NaX. This large difference (*circa* 200 K) impacts the operating conditions of a heat storage device. Hydrophobic materials such as HS-FAU can be used at low-temperature conditions, *e.g.*, in the 300-350 K range. Simultaneously, hydrophilic adsorbents can operate in high-temperature processes with desorption temperatures over 550 K. This suggests the possibility of tuning the Si/Al ratio to maximize the efficiency of adsorbate-fluid working pairs for given operational conditions.

We calculated characteristic curves, adsorption enthalpy, and storage densities of the working pairs using a thermodynamical model based on the theory of adsorption of Dubinin-Polanyi. The choice of the operating conditions for each adsorbent-fluid working pair is crucial. This is a limiting factor for the performance of materials or working fluids. The thermodynamical model provides insights on the performance of a heat storage device by only combining adsorption data with some physicochemical properties of the fluids. These are the density, the saturation pressure, and the enthalpy of vaporization in a range of operational temperatures and pressures. These properties can be obtained from experiments, but also from molecular simulation. This could be useful for the screening of adsorbent-fluid working pairs oriented to energy storage applications.

The energy released upon heating and cooling a fluid is higher for water than for methanol. We found storage densities of the water-zeolite pairs higher than 1100 kJ/kg, while for methanol-zeolite pairs were about 350 kJ/kg. The water/methanol ratio of storage densities is related to the ratio of their adsorption loading. The highest values of water uptake are due to both, a strongest hydrogen bond network and the access of water to the sodalite cages of the FAU zeolites. The agreement found between experiments and simulation allows the use of GCMC simulation for other operational conditions and provides a comprehensive overview of the performance of the working pairs for energy storage.

## AUTHOR INFORMATION
Jose Manuel Vicent-Luna[*]
* E-mail: j.vicent.luna@tue.nl

Sofía Calero[*]
* E-mail: s.calero@tue.nl

**Notes**
The Authors declare no competing financial interest

## ACKNOWLEDGMENTS
This work was supported by ERC ZEOSEP Ref.: 779792, MINECO CTQ2016-80206-P and CTQ2017-95-173-EXP. A.S. obtained financial resources as part of financing the doctoral scholarship from the National Science Center, Poland, grant no. 2018/28/T/ST5/00274. We thank C3UPO for the HPC support.

# SUPPORTING INFORMATION

## of the paper entitled

## On the use of Water and Methanol with Zeolites for Heat Transfer

Rafael M. Madero-Castro[a], Azahara Luna-Triguero[b,c], Andrzej Sławek[d,e], José Manuel Vicent-Luna[f]\*, and Sofia Calero[a,f]\*

[a]Department of Physical, Chemical, and Natural Systems, Universidad Pablo de Olavide. Ctra. Utrera km. 1. ES-41013 Seville, Spain.
[b]Energy Technology, Department of Mechanical Engineering, Eindhoven University of Technology, P.O. Box 513, 5600 MB Eindhoven, The Netherlands.
[c]Eindhoven Institute for Renewable Energy Systems (EIRES), Eindhoven University of Technology, PO Box 513, Eindhoven 5600 MB, the Netherlands
[d]Academic Centre for Materials and Nanotechnology, AGH University of Science and Technology, Kawiory 30, 30-055 Kraków, Poland.
[e]Faculty of Chemistry, Jagiellonian University, Gronostajowa 2, 30-387 Kraków, Poland.
[f]Materials Simulation and Modelling, Department of Applied Physics, Eindhoven University of Technology, P.O. Box 513, 5600MB Eindhoven, The Netherlands.

**Corresponding Authors**

\* José Manuel Vicent-Luna, e-mail: j.vicent.luna@tue.nl;
\* Sofía Calero, e-mail: s.calero@tue.nl

**Number of pages: 6      Number of tables: 2      Number of figures: 1**

**Outline**

**Section S1.** Structural model of zeolites.

All zeolites were generated following the same procedure regardless the Si/Al ratio. The unit cell of the pure silica FAU contains 192 Si atoms. We substitute some Si atoms by Al atoms to reproduce the experimental chemical composition (Si/Al ratio of 100, 2.61, and 1.06, HS-FAU, NaY, and NaX, respectively). HS-FAU, NaY, and NaX contain 2, 56, and 88 Al atoms, respectively.

We generated the structures following the methodology described in previous works [1,2]. We started from the crystallographic positions of the pure silica zeolite from the International Zeolite Association (IZA) database [3] to construct the aluminosilicates. For each structure, we created a set of 50 configurations by randomly substituting some silicon atoms by aluminum atoms within the constraint of Löwenstein's rule and selected the most energetically favorable configuration. Then, we compensated the net negative charge of the adsorbents by placing sodium extra-framework cations in the most probable crystallographic positions reported in the literature. A detailed description of these extra-framework cations is given in references [4-6]. Once we added the extra-framework cations to their preferential location, we optimized the structures with energy minimization simulations using Baker's [7] method and a full-flexible core-shell potential.[8,9]

**Section S2.** Parameterization of methanol-zeolite interactions.

Interactions parameters between the molecules of water and the HS-FAU and NaX zeolites were developed in our previous work [10] using experimental adsorption isobars as reference data. In this work, we also computed the adsorption isobar of water in NaY, showing that the water-zeolite interactions are transferable in the whole range of Si/Al substitutions. Here we followed a similar procedure to obtain the methanol-zeolite interactions. Starting from the cross-term Lennard-Jones parameters for each pseudo atom of the methanol-zeolite pairs, we iteratively modify the ε and σ parameters, creating a matrix of values smaller and larger than the initials. The partial charges for the adsorbates and zeolites are kept fixed and given in Table S1. For each set of parameters, we computed five values of an adsorption isobar from the low to the high coverage regime. We first compare with experimental data for NaX to narrow the search of adequate parameters to reproduce the adsorption in the zeolite with the highest content of extra-framework cations. Then, we compare with the measured data for HS-FAU and finally for NaY. We repeated the process until we found reasonable agreement between experiments and simulations using the same set of Lennard-Jones parameters regardless the Si/Al ratio. The optimal values are provided in Table S2.

**Table S1.** Charges of each pseudo-atom for the adsorbent and adsorbates. $O_{zeo}$-Si and $O_{zeo}$-Al are the oxygen atoms bridging to silicon and aluminum atoms, respectively.

| Molecule | Atom | q[$e$] |
|---|---|---|
| water[a] | $O_w$ | -0.8476 |
| | $H_w$ | 0.4238 |
| methanol[b] | $CH_3$ | 0.265 |
| | $O_{alc}$ | -0.7 |
| | $H_{alc}$ | 0.435 |
| zeolite[c] | $Si_{zeo}$ | 2.05 |
| | $O_{zeo-Si}$ | -1.025 |
| | $Al_{zeo}$ | 1.75 |
| | $O_{zeo-Al}$ | -1.2 |
| | Na | 1.0 |

[a] reported in reference [11], [b] reported in reference [12], and [c] reported in reference [13],

**Table S2.** Lennard-Jones parameters to describe the interactions between the zeolite and the water and methanol molecules.

| Molecule | Pair interaction | $\varepsilon/k_B$ [K] | $\sigma$ [Å] |
|---|---|---|---|
| **water**[a] | $O_w$-$O_{zeol}$ | 80 | 3.3 |
|  | $O_w$-$Na^+$ | 50 | 3.3 |
| **methanol**[b] | $CH_3$-$O_{zeol}$ | 80.0 | 3.8 |
|  | $O_{alc}$-$O_{zeol}$ | 70.0 | 3.8 |
|  | $CH_3$-$Na^+$ | 80.0 | 3.6 |
|  | $O_{alc}$-$Na^+$ | 80.0 | 3.2 |

[a] Taken from ref [10]. [b] This work.

**Section S3.** Thermodynamical model.

We used the mathematical model based on the Dubinin-Polanyi theory [14,15] to obtain the energy storage properties of the zeolite-fluid working pairs. We first convert the adsorption isobars into their corresponding characteristic curves. The characteristic curve relates the volumetric uptake $W$ (volume of fluid adsorbed in the micropores [ml/g]) and the adsorption potential $A$ [kJ/mol].

$$A = RT \ln p_{sat}/p \quad (1)$$

$$W = q(p,T)/\rho(T) \quad (2)$$

where $p_{sat}$, is the saturation pressure of the working fluid, $q(p,T)$, the loading of adsorbed fluid per mass of adsorbent, [g/g], and $\rho(T)$, the density of fluid confined within the micropores [g/ml]. We use the Peng Robinson equation of state to calculate the saturation pressure of each fluid.[16] We obtained the loading of fluid from QE-TPDA experiments and GCMC simulation. We used the model of Hauer to obtain the density of confined fluids within the micropores.[17,18] This model gives a linear relationship between the density of a fluid confined within the pores of an adsorbent and the operational temperature:

$$\rho(T) = \rho_0(T_0) \cdot [1 - \alpha_T(T - T_0)] \quad (3)$$

where $\rho_0$ is the free liquid density at the reference $T_0$ (283.15 K for water [15] and 298 K for metanol [19]). $\alpha_T$ is the free liquid thermal expansion coefficient of each working fluid at the reference temperature and 100 MPa [18,19] (3.871·10$^{-4}$ K$^{-1}$ for water and 8.026·10$^{-4}$ K$^{-1}$ for methanol).

The Dubinin-Polanyi theory also allows determining the adsorption enthalpy, which is defined as:

$$\Delta H = \Delta H_{vap} + A - T\Delta S \quad (4)$$

where $\Delta H_{vap}$ is the enthalpy of vaporization, $A$ is the adsorption potential (Gibbs free energy) and $\Delta S$ is the entropy variation,[20] calculated as:

$$\Delta S = \alpha_{Ads} W \frac{\partial A}{\partial W}\bigg|_T \quad (5)$$

where $\alpha_{Ads}$ is the thermal expansion coefficient of the fluid in the adsorbed phase, obtained from the density model. Finally, we calculated the thermochemical storage density by integrating the enthalpy curves within the selected adsorption and desorption temperatures:

$$SD = \int_{q(T_{ads})}^{q(T_{des})} \Delta H(q) dq \quad (6)$$

In summary, the mathematical model based on the Dubinin-Polanyi theory allows obtaining the storages densities of adsorbent-fluid working pairs, just from an adsorption isotherm or isobar and some physicochemical properties of the fluids. These properties are the enthalpy of vaporization, bulk liquid density, thermal expansion coefficient, and saturation pressure.

Another advantage of the characteristic curve is that it can be reverted to obtain the adsorption isobars or isotherms at different conditions. In addition to the adsorption isobars, we also computed the adsorption isotherms to check the validity of the Dubinin-Polanyi theory. Figure S1 shows the adsorption isotherms of both working fluids in the three zeolites.

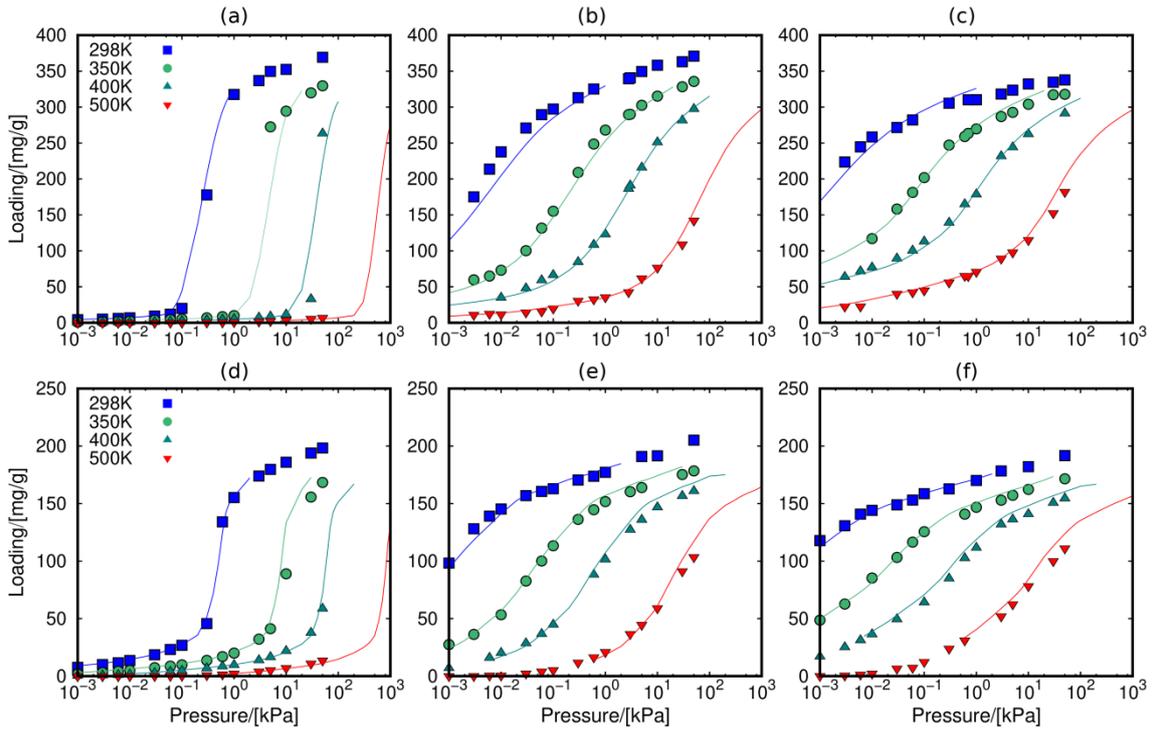

**Figure S1**. Adsorption isotherms at 298, 350, 400, and 500 K of water (top) and methanol (bottom) in HS-FAU (a, d), NaY (b, e), and NaX (c, f), respectively. Closed symbols represent computed values with GCMC simulations and lines are predicted isotherms obtained from the characteristic curves using the thermodynamical model of Dubinin-Polanyi.